\documentstyle[preprint,pra,eqsecnum,aps]{revtex}
\tightenlines
\begin{document}
\draft

\title{{\Large \bf SQUEEZED STATES AS REPRESENTATIONS OF SYMPLECTIC GROUPS}}

\author{D. Han\footnote{electronic mail: han@trmm.gsfc.nasa.gov}}
\address{National Aeronautics and Space Administration, Goddard Space
Flight Center, Code 910.1, Greenbelt, Maryland 20771}

\author{Y. S. Kim\footnote{electronic mail: kim@umdhep.umd.edu}}
\address{Department of Physics, University of Maryland, College Park,
Maryland 20742}

\maketitle

\begin{abstract}
It is shown that the $SU(1,1)$-like and $SU(2)$-like two-photon
coherent states can be combined to form a $O(3,2)$-like two-photon
states.  Since the $O(3,2)$ group has many subgroups, there are
also many new interesting new coherent and squeezed two-photon states.
Among them is the two-photon sheared state whose symmetry property
is like that for the two-dimensional Euclidean group.  There are now
two-phonon coherent states which may exhibit symmetries not yet observed
for photons, including sheared states.  Let us note that both $SU(1,1)$
and $S(3,2)$ are isomorphic to the symplectic groups $Sp(2)$ and
$Sp(4)$ respectively, and that symplectic transformations consist of
rotations and squeeze transformations.
\end{abstract}

\section{Introduction}\label{intro}
We are quite familiar with rotations, but not with squeeze
transformations.  Yet, squeeze transformations are everywhere in
physics, including special relativity, classical mechanics, and
quantum mechanics.  If we enlarge the group of rotations by including
squeeze transformations, it becomes a symplectic group.  For this
reason, most of the groups in physics are symplectic groups or
subgroups of the symplectic group, and quantum optics is not an
exception.  We have squeezed states of light!

Let us first look at the two two-mode states discussed in detail by
Yurke, McCall and Klauder~\cite{ymk86}.  One of the states is generated
by the operators satisfying the commutation relations for the $SU(2)$
group, and the other state generated by those satisfying the $SU(1,1)$
algebra or the algebra of the $Sp(2)$ symplectic
group~\cite{hkn88,knp91}.  If we combine these two states by
constructing a set of closed commutation relations starting from the
generators of both states, we end up with ten generators satisfying the
commutation relations for the $O(3,2)$
Lorentz group~\cite{knp91,bishop88,hkn90}.  This $O(3,2)$-like
two-oscillator state was considered by Dirac in 1963~\cite{dir63}.

Although quantum optics has been the primary beneficiary of the
symmetries exhibited by two-photon states in recent years, we should
note that squeezed states are not restricted to optical sciences.  It
was noted that the $O(3,2)$ symmetry is relevant to Bogoliubov
transformations in superconductivity~\cite{kibir88}.  There are now
two-phonon coherent states~\cite{garret97,rojo97} possessing symmetry
properties not observed in photon states.  It is quite possible that
they symmetries derivable from the $E(2)$-like subgroups of the
$O(3,2)$-like two-oscillator system.

In Secs~\ref{squeeze}, \ref{subg}, and \ref{shear} we discuss how
the $O(3,2)$ states are constructed from the $SU(2)$ and $SU(1,1)$
states, the $E(2)$-like little group of $O(3,2)$, and the $E(2)$-like
sheared states, respectively.  Our discussion of the sheared state
is largely based on the 1992 paper by Kim and Yeh~\cite{ky92}.

\section{Squeezed States}\label{squeeze}

It is a well-established tradition that coherent and squeezed states
as well as their variants are represented by three-parameter groups.
The symmetry of coherent state is a representation of the
three-parameter Heisenberg group.

The one-mode squeezed state is representation of the $U(1,1)$ group
generated by
\begin{equation}
{1\over 2}\left(\hat{a}^{\dag}\hat{a} +
                       \hat{a}\hat{a}^{\dag}\right), \qquad
{1\over 2}\left(\hat{a}^{\dag}\hat{a}^{\dag} +
                       \hat{a}\hat{a}\right), \qquad
{i\over 2}\left(\hat{a}^{\dag}\hat{a}^{\dag} - \hat{a}\hat{a}\right) .
\end{equation}
which is locally isomorphic to $Sp(2)$, generated by
\begin{equation}
{1\over 2}\pmatrix{0 & -i \cr i & 0} , \qquad
{1\over 2}\pmatrix{i & 0 \cr 0 & -i} , \qquad
{1\over 2}\pmatrix{0 & i \cr i & 0} ,
\end{equation}
which satisfy the same set of commutation relations as that for the
generators of the $SU(1,1)$ group.

For two-mode squeezed states, the $SU(2)$ and $SU(1,1)$ symmetries have
been discussed extensively in the literature~\cite{ymk86,knp91,cav85}.
Those are generated respectively
\begin{equation}\label{su2}
\hat{J}_{1} = {1\over 2}\pmatrix{\hat{a}^{\dag}_{1}\hat{a}_{1} -
\hat{a}^{\dag}_{2}\hat{a}_{2}} ,  \quad
\hat{J}_{2} = {1\over 2}\pmatrix{\hat{a}^{\dag}_{1}\hat{a}_{2} +
\hat{a}^{\dag}_{2}\hat{a}_{1}} , \quad
\hat{J}_{3} = {1\over 2i}\pmatrix{\hat{a}^{\dag}_{1}\hat{a}_{2} -
\hat{a}^{\dag}_{2}\hat{a}_{1}}
\end{equation}
and
\begin{equation}\label{su11}
\hat{J}_{o} = {1\over 2}\pmatrix{\hat{a}^{\dag}_{1}\hat{a}_{1}
 + \hat{a}_{2}\hat{a}^{\dag}_{2}} ,  \quad
\hat{K}_{1} = {1\over 2}\pmatrix{\hat{a}^{\dag}_{1}\hat{a}^{\dag}_{2}
+ \hat{a}_{1}\hat{a}_{2}} , \quad
\hat{Q}_{1} = {i\over 2}\pmatrix{\hat{a}^{\dag}_{1}\hat{a}^{\dag}_{2}
- \hat{a}_{1}\hat{a}_{2}} .
\end{equation}
If we take commutation relations of the six operators given in
Eq.(\ref{su2}) and Eq.(\ref{su11}), they generate the following four
additional generators.
\begin{eqnarray}
\hat{K}_{2} &=&-{1\over 4}\pmatrix{\hat{a}^{\dag}_{1}\hat{a}^{\dag}_{1}
+ \hat{a}_{1}\hat{a}_{1} - \hat{a}^{\dag}_{2}\hat{a}^{\dag}_{2} -
\hat{a}_{2}\hat{a}_{2}} , \quad
\hat{K}_{3} = {i\over 4}\pmatrix{\hat{a}^{\dag}_{1}\hat{a}^{\dag}_{1}
- \hat{a}_{1}\hat{a}_{1} + \hat{a}^{\dag}_{2}\hat{a}^{\dag}_{2} -
\hat{a}_{2}\hat{a}_{2}} , \nonumber \\[2ex]
\hat{Q}_{2} &=&-{i\over 4}\pmatrix{\hat{a}^{\dag}_{1}\hat{a}^{\dag}_{1}
- \hat{a}_{1}\hat{a}_{1} - \hat{a}^{\dag}_{2}\hat{a}^{\dag}_{2} +
\hat{a}_{2}\hat{a}_{2}} , \quad
\hat{Q}_{3} = -{1\over 4}\pmatrix{\hat{a}^{\dag}_{1}\hat{a}^{\dag}_{1}
+ \hat{a}_{1}\hat{a}_{1} + \hat{a}^{\dag}_{2}\hat{a}^{\dag}_{2} +
\hat{a}_{2}\hat{a}_{2}} .
\end{eqnarray}
There are ten linearly independent generators.  In terms of the Pauli
matrices, they are~\cite{hkn95jm}
\begin{eqnarray}\label{pauli}
&{}& J_{1} = {i\over 2}
     \pmatrix{0 & \sigma_{3} \cr -\sigma_{3} & 0} , \quad
     J_{2} = {i\over 2}\pmatrix{0 & \sigma_{1} \cr
      -\sigma_{1} & 0}, \quad
     J_{3} = {1\over 2}
     \pmatrix{\sigma_{2} & 0 \cr 0 & \sigma_{2}}, \quad
     J_{0} = {i\over 2} \pmatrix{0 & I \cr -I & 0} , \nonumber \\[2ex]
&{}& K_{1} = {-i\over 2}
     \pmatrix{0 & \sigma_{1} \cr \sigma_{1} & 0}, \quad
     K_{2} = {i\over 2}\pmatrix{0 & \sigma_{3} \cr
     \sigma_{3} & 0},\quad
     K_{3} = {i\over 2} \pmatrix{I & 0 \cr 0 & -I} , \nonumber\\[2ex]
&{}& Q_{1} = {i\over 2}
     \pmatrix{\sigma_{1} & 0 \cr 0 &-\sigma_{1}} , \quad
     Q_{2} = {-i\over 2}\pmatrix{\sigma_{3} & 0 \cr
     0 & -\sigma_{3}}, \quad
     Q_{3} = {i\over 2}\pmatrix{0 & I \cr I & 0} .
\end{eqnarray}
These generators satisfy the commutation relations:
\begin{eqnarray}
&{}& [J_{i}, J_{j}] = i\epsilon_{ijk} J_{k} , \quad
     [J_{i}, K_{j}] = i\epsilon_{ijk} K_{k} , \quad
     [J_{i}, Q_{j}] = i\epsilon_{ijk} Q_{k} , \quad
     [K_{i}, Q_{j}] = i\delta_{ij} J_{0} ,
     \nonumber \\[2ex]
&{}& [K_{i}, K_{j}] = [Q_{i}, Q_{j}] =
     -i\epsilon_{ijk} J_{k} , \quad
     [J_{i}, J_{o}] = 0 ,  \quad
     [K_{i}, J_{0}] = iQ_{i} , \quad
     [Q_{i}, J_{0}] = -iK_{i} .
\end{eqnarray}
Indeed, these are the generators of the group $Sp(4)$ which is locally
isomorphic to the (3~+~2)-dimensional deSitter group~\cite{hkn90,dir63}.

\section{Three-parameter Subgroups of Sp(4)}\label{subg}
The group $Sp(4)$ has many subgroups, and three-parameter subgroups
are particularly useful in studying two-mode squeezed
states~\cite{ymk86,hkn90}.  The most obvious subgroup is the
$SU(2)$-like group generated by $J_{1}, J_{2}$, and $J_{3}$.
Another subgroup frequently discussed in the literature is the
$SU(1,1)$-like
group generated by $J_{o}, K_{3}, Q_{3}$.  From the local isomorphism
between $Sp(4)$
and $O(3,2)$, it is not difficult to find all possible $SU(1,1)$-like
subgroups.

Indeed, the above-mentioned $SU(1,1)$-like subgroup is unitarily
equivalent to those generated by $J_{o}, K_{1}, Q_{1}$ and by $J_{0}, K_{2},
Q_{2}$.  There is also an $SU(1,1)$-like subgroup generated by $K_{1},
K_{2}, J_{3}$, which is unitarily
equivalent to $Q_{1}, Q_{2}$, and $J_{3}$.  These are then unitarily
equivalent to the subgroups generated by $K_{2}, K_{3}, J_{1}$,
by $K_{3}, K_{1}, J_{2}$, by $Q_{2}, Q_{3}, J_{1}$, and by $Q_{3},
Q_{1}, J_{2}$.  It is known from the Lorentz group that the signs of
$K_{i}$ and $Q_{i}$ can be changed.

The purpose of this paper is to discuss additional three-parameter
subgroups.  It is known that the $O(3,1)$ Lorentz group has a number of
$E(2)$-like subgroups.  Since there are two $O(3,1)$-like subgroups in
the $O(3,2)$ deSitter group, the group $Sp(4)$ should contain a number
of $E(2)$-like subgroups~\cite{hkn95jm}.  If we define
\begin{equation}
F_{1} = K_{1} - J_{2} , \qquad F_{2} = K_{2} + J_{1} ,
\end{equation}
and
\begin{equation}
G_{1} = Q_{1} - J_{2} , \qquad G_{2} = Q_{2} + J_{1} ,
\end{equation}
then the resulting commutation relations are
\begin{equation}\label{e2com}
[F_{1}, F_{2}] = 0 , \qquad [J_{3}, F_{1}] = - iF_{2} , \qquad
[J_{3}, F_{2}] =  iF_{1} ,
\end{equation}
and similar relations for $G_{1}, G_{2}$ and $J_{3}$. This set
of commutation relations is the same as for the generators of the
$E(2)$ group.  The subgroup generated by
$G_{1}, G_{2}$ and $J_{3}$ is unitarily equivalent to that generated
by $F_{1}, F_{2}$ and $J_{3}$.  There are four-additional subgroups
generated by $F_{2}, F_{3}, J_{1}$, by $F_{3}, F_{1,} J_{2}$, by
$G_{2}, G_{3}, J_{1}$, and by $G_{3}, G_{1}, J_{2}$.  They are all
unitarily equivalent.  Thus we can study them all by studying one of
them.

\section{Construction of Sheared States}\label{shear}
Let us define the word ``shear.''  In the $xp$ plane, we can consider
linear transformations of the type
\begin{equation}\label{xpshear}
\pmatrix{x'\cr p'} = \pmatrix{1&\alpha \cr0&1} \pmatrix{x\cr p} ,
\end{equation}
Under this transformation, the $x$ coordinate undergoes a translation
proportional to $p$ while the $p$ variable remains unchanged.
\begin{equation}\label{pxshear}
\pmatrix{x'\cr p'} = \pmatrix{1 & 0 \cr \alpha & 1} \pmatrix{x \cr p} .
\end{equation}
For a given function $f(x,p)$, the shears of Eq.(\ref{xpshear}) and
Eq.(\ref{pxshear}) lead to $f(x - \alpha p, p)$ and $f(x, p - \alpha x)$
respectively.

With this preparation, we continue our discussion of the $E(2)$-like
subgroups discussed introduced in Sec.~\ref{shear}.  In the four-by-four
matrix representation, the rotation generator $J_{3}$ is given in
Eq.(\ref{pauli}).  The generators $F_{1}$ and $F_{2}$ take the matrix
form.
\begin{equation}
F_{1} = -i\pmatrix{0 & \sigma_{1} \cr 0 & 0} ,\qquad
F_{2} = i\pmatrix{0 & \sigma_{3} \cr 0 & 0} .
\end{equation}
and the transformation matrix become
\begin{equation}
S(\alpha,\beta) = \exp{\left( -i\alpha F_{1} -i\beta F_{2} \right)}
= \pmatrix{I & -\alpha\sigma_{1} + \beta\sigma_{3}
\cr 0 & I} .
\end{equation}

We can now write the above operator in terms of creation and
annihilation operators.  Indeed, if we compare the expressions given
in Eq.(\ref{su2}), Eq.(\ref{su11}) and Eq.(\ref{pauli}),
\begin{equation}
\hat{S}(\alpha,\beta) = \exp{\left({\alpha\over 4}
\left\{\left(\hat{a}^{\dag}_{1} - \hat{a}_{1}\right)^{2} -
\left(\hat{a}^{\dag}_{2} - \hat{a}_{2}\right)^{2}\right\}
+ {\beta\over 2}\left(\hat{a}^{\dag}_{1} - \hat{a}_{1}\right)
\left(\hat{a}^{\dag}_{2} - \hat{a}_{2}\right)\right)} .
\end{equation}
The exponent of this expression takes the quadratic form
\begin{equation}
{\alpha\over 4}\left(\hat{a}^{\dag2}_{1} +
\hat{a}^{2}_{1} - 2\hat{a}^{\dag}_{1}\hat{a}_{1}\right) -
{\alpha\over 4}\left(\hat{a}^{\dag2}_{2} + \hat{a}^{2}_{2} -
2\hat{a}^{\dag}_{2}\hat{a}_{2}\right) +
{\beta\over 2}\left(\hat{a}^{\dag}_{1}\hat{a}^{\dag}_{2} +
\hat{a}_{1}\hat{a}_{2} - \hat{a}^{\dag}_{2}\hat{a}_{1} -
\hat{a}^{\dag}_{1}\hat{a}_{2}\right) .
\end{equation}
This is normal-ordered, but the shear operator $\hat{S}(\alpha,\beta)$
is not.  However, there are theorems in the literature which allow us
to write the $\hat{S}$ operator in a normal-ordered form~\cite{agar70}.
Indeed, $\hat{S}(\alpha, \beta)$ can be written as
\begin{eqnarray}
\lefteqn{\hat{S}(\alpha,\beta) = \lambda :
\exp\left\{\xi\pmatrix{\hat{a}^{\dag2}_{1} + \hat{a}^{2}_{1} -
2\hat{a}^{\dag}_{1}\hat{a}_{1} }
+ \eta\pmatrix{\hat{a}^{\dag2}_{2} + \hat{a}^{2}_{2} -
2\hat{a}^{\dag}_{2}\hat{a}_{2}} \right.}  \nonumber \\[2ex]
\mbox{} & \mbox{} & \mbox{} \hspace{70mm}
\left. + \zeta\pmatrix{\hat{a}^{\dag}_{1}\hat{a}^{\dag}_{2} +
\hat{a}_{1}\hat{a}_{2} - \hat{a}^{\dag}_{1}\hat{a}_{2} -
\hat{a}^{\dag}_{2}\hat{a}_{1}} \right\}: ,
\end{eqnarray}
with
\begin{equation}\label{coeffs}
\lambda = {2 \over \sqrt{\alpha ^{2} + \beta^{2} + 4}} , \quad
\xi = \frac{\alpha^{2} + \beta^{2} - 2i\alpha}
      {2\pmatrix{\alpha^{2} + \beta^{2} + 4}} , \quad
\eta = \frac{\alpha^{2} + \beta^{2} + 2i\alpha}
   {2\pmatrix{\alpha^{2} + \beta^{2} + 4}} ,  \qquad
   \zeta = \frac{-2i\beta}{\alpha^{2} + \beta^{2} + 4} .
\end{equation}
If this operator is applied to the vacuum state, the annihilation operators
deleted.  The result is
\begin{equation}
\hat{S}(\alpha,\beta)|0,0> = \hat{T}(\alpha,\beta )|0,0> ,
\end{equation}
where
\begin{equation}
\hat{T}(\alpha,\beta)= \lambda
\exp{\pmatrix{\xi \hat{a}^{\dag2}_{1} + \eta \hat{a}^{\dag2}_{2}
+ \zeta \hat{a}^{\dag}_{1}\hat{a}^{\dag}_{2}}} .
\end{equation}
This operator can now be decomposed into
\begin{equation}
\hat{T}(\alpha,\beta)= \lambda \exp{\pmatrix{\xi \hat{a}^{\dag2}_{1}}}
\exp{\pmatrix{\eta \hat{a}^{\dag2}_{2}}}
\exp{\pmatrix{\zeta \hat{a}^{\dag}_{1}\hat{a}^{\dag}_{2}}} .
\end{equation}

If this operator is acted on the vacuum state,
\begin{equation}\label{series}
|\alpha,\beta> = \hat{T}(\alpha, \beta)|0> =
\sum^{}_{k,j} C_{kj} |k,j> ,
\end{equation}
where $k$ and $j$ are the Fock-space indices for the first and second
photons respectively.
Using the form of Eq.(\ref{coeffs}), it is possible to calculate the
coefficient $C_{kj}$.  The results are
\begin{eqnarray}
&{}&   C_{2m+1,2n} = C_{2m,2n+1} = 0 ,  \nonumber \\[2ex]
&{}&  C_{2m,2n} = \lambda \sqrt{(2m)!(2n)!} \hspace{3mm}
\sum^{Min(m,n)}_{k=0}\frac{\zeta^{2k}\xi^{m-k}\eta^{n-k}}
{(2k)!(m-k)!(n-k)!} ,  \nonumber \\[2ex]
&{}&  C_{2m+1,2n+1} = \lambda\sqrt{(2m + 1)!(2n + 1)!} \hspace{3mm}
\sum^{Min(m,n)}_{k=0}\frac{\zeta^{2k+1}\xi^{m-k}\eta^{n-k}}
{(2k+1)!(m-k)!(n-k)!} .
\end{eqnarray}
When $\beta = 0, C_{2m+1,2n+1} = 0$, and the series of Eq.(\ref{series})
becomes a separable expansion.

The distribution $|C_{k,j}|^{2}$ of course depends on $k$ and $j$
which are the numbers for the first and second photons.
These coefficients are normalized.  The average value of the number
of the first and second photons are
\begin{equation}
<N_{1}> = \sum^{\infty}_{k=0} k |C_{k,j}|^{2} , \qquad
<N_{2}> = \sum^{\infty}_{j=0} j |C_{k,j}|^{2} .
\end{equation}
Similarly,
\begin{equation}
<N_{1}N_{2}> = \sum^{\infty}_{k=0} kj |C_{k,j}|^{2} ,
\end{equation}
and
\begin{equation}
<N^{2}_{1}> = \sum^{\infty }_{k=0} k^{2}|C_{k,j}|^{2} , \qquad
<N^{2}_{2}> = \sum^{\infty }_{j=0} j^{2}|C_{k,j}|^{2} .
\end{equation}
The computation of these numbers in terms of the coefficients given in
Eq.(\ref{coeffs}) is possible.  On the other hand, it is not clear
whether this computation will lead to a closed analytical form for each
of the above quantities.  The method of Wigner function will make these
calculations relatively simple~\cite{knp91,ky92}.

In this report, we have given only a sample calculation.  We have not
made any attempt to derive the numbers which can be measured in
laboratories.  This can be done when there are more concrete evidence
for sheared states of photons or phonons~\cite{garret97,rojo97}.

\end{document}